\documentclass[10pt,letterpaper]{article}
\usepackage[top=0.85in,left=2.75in,footskip=0.75in,marginparwidth=2in]{geometry}

\usepackage[utf8]{inputenc}

\usepackage{cite}

\usepackage{nameref,hyperref}

\usepackage[right]{lineno}

\usepackage{microtype}
\DisableLigatures[f]{encoding = *, family = * }

\raggedright
\usepackage{changepage}

\usepackage[aboveskip=1pt,labelfont=bf,labelsep=period,singlelinecheck=off]{caption}

\makeatletter
\renewcommand{\@biblabel}[1]{\quad#1.}
\makeatother

\usepackage{lastpage,fancyhdr,graphicx}
\usepackage{epstopdf}
\fancyheadoffset[L]{2.25in}
\fancyfootoffset[L]{2.25in}

\usepackage{color}

\definecolor{Gray}{gray}{.25}

\usepackage{graphicx}

\usepackage{sidecap}

\usepackage{wrapfig}
\usepackage[pscoord]{eso-pic}
\usepackage[fulladjust]{marginnote}
\reversemarginpar

\begin{document}
\vspace*{0.35in}

\begin{flushleft}
{\Large
\textbf\newline{Revisiting the Global Minimum Structure of the Pt$_5$V Cluster}
}
\newline
\\
P. L. Rodr\'iguez-Kessler\textsuperscript{1,*}
\\
\bigskip
\bf{1} Centro de Investigaciones en \'Optica A.C., Loma del Bosque 115, Lomas del Campestre , Leon, 37150, Guanajuato, Mexico
\bigskip
*plkessler@cio.mx

\end{flushleft}

\justifying

\section*{Abstract}
 In this work, the most stable structures of Pt$_5$V clusters are investigated using the successive growth algorithm (SCG) in conjunction with density functional theory (DFT) calculations. The resulting structures are evaluated by various functionals, including GGA (PBE), meta-GGA (TPSS) and hybrid (B3PW91, PBE0, PBEh-3c, M06-L), in conjunction with the Def2TZVP basis set. The results based on these functionals show that two different structures compete for the global minimum. The structural and electronic properties of the two most stable structures are discussed.

\section*{Introduction}

Recently, a significant number of theoretical and experimental studies have been devoted to understanding the structure-performance properties of both charged and neutral clusters.\cite{NI2024416183,SHI2024114738,Tian2024,Li2024,Tian2024} In particular, platinum-based nanoclusters have been widely investigated due to their extensive use in heterogeneous catalysis. Research in platinum materials led to the evaluation of catalytic active sites, enabling further strategies toward more efficient catalytic materials. The high cost of platinum (Pt) and other platinum group metals has made the study of heterogeneous nanoalloy systems increasingly important. In particular, Pt alloyed with non-precious metals (M) offer intrinsic cost reductions and also shows an increase in the catalytic activities.\cite{Stamenkovic2007ER, XIONG2002898} For pure platinum clusters, the structures have been conclusively studied for small sizes, while for atomicity 16 and onward, the global minimum structures are still under debate.\cite{10.1063/1.4935566,D2CP05188E,D3CP04455F,10.1063/1674-0068/cjcp2309083} On the other hand, studies on doped Pt clusters with early transition metals are limited.\cite{10.1063/1.2839437,D1CP05410D,Arunachalam2022,RODRIGUEZKESSLER2020155897,D1CP00379H} This is not a trivial task, as the doped clusters require extensive exploration of their structures, chemical ordering, and spin.\cite{RODRIGUEZKESSLER201534,Ponce-Tadeo2016,PhysRevB.92.125442,doi:10.1021/acs.jpca.6b00224} In the literature, Jennings and Johnston revealed the most stable structures of Pt$_{x-y}$M$_{y}$ (with x=2-6, y=1,2 and M=Ti, V) clusters and found that the spin states have a negligible effect on the structures of the clusters.\cite{JENNINGS201391} In order to confirm the lowest energy structure of these clusters, in this preprint, we test the case of Pt$_5$V clusters using various functionals. The results show that the global minimum structure exhibits functional dependency behavior. The vibrational and electronic properties of representative Pt$_5$V clusters are discussed.

\section*{Materials and Methods}

The calculations in this study utilize density functional theory (DFT) as implemented in the Orca quantum chemistry package. \cite{10.1063/5.0004608}. The Exchange and correlation energies are addressed using various functionals, including the GGA (PBE),\cite{PhysRevLett.77.3865} meta-GGA (TPSS)\cite{PhysRevLett.91.146401} and hybrid (B3PW91, PBE0, PBEh-3c, M06-L)\cite{10.1063/1.464913,10.1063/1.4927476,10.1063/1.478522,10.1063/1.2370993} in conjunction with the Def2TZVP basis set.\cite{B508541A} These functionals were selected according to a benchmark calculation on Pt clusters, which showed a good aggrement with the experimental data.\cite{D3CP04455F} Atomic positions are self-consistently relaxed through a Quasi-Newton method employing the BFGS algorithm. The SCF convergence criteria for geometry optimizations are achieved when the total energy difference is smaller than 10$^{-8}$ au, by using the TightSCF keyword in the input. The  Van  der  Waals  interactions  are  included in the exchange-correlation functionals with empirical dispersion corrections of Grimme DFT-D3(BJ). {\cite{https://doi.org/10.1002/jcc.21759}} The structure search was conducted using a biased search structure as implemented in the SCG method.\cite{D0CP06179D}


\section*{Results}

\begin{figure*}[ht!]
\Large
  \vspace{.5cm}
  \begin{tabular}{l}
\includegraphics[scale=0.55]{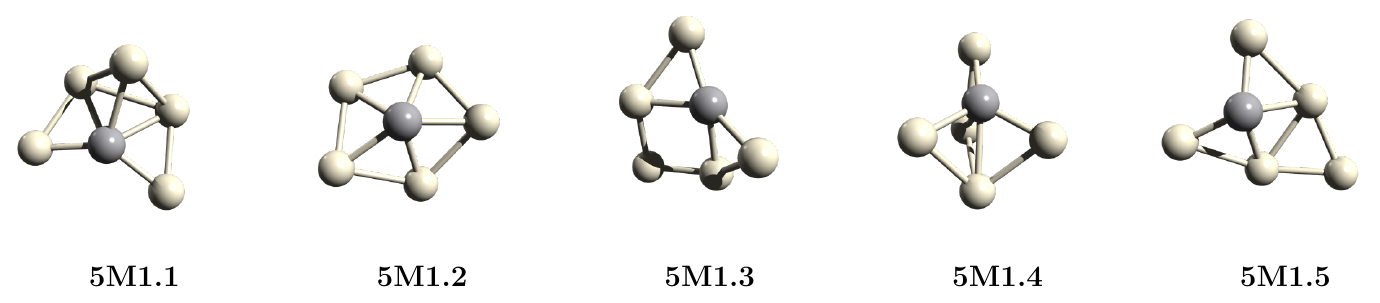} \\
\end{tabular}
       \caption{\label{figure_struc}Lowest energy structures of Pt$_5$V clusters. The structures are labeled with the {\bf nMm.y} notation, where {\bf n}:{\bf m} stands for the number of Pt:V atoms, {\bf M} denotes the doped system, and {\bf y} refers to the isomer number in increasing energy ordering.}
\end{figure*}

 By exploring the structures of Pt$_5$V clusters, we identified five representative structure isomers, as shown in Figure~\ref{figure_struc}. Previous comparisons of DFT methods with experimental data suggest that hybrid-GGA and meta-GGA functionals are good choices for evaluating the structures of platinum clusters.\cite{D3CP04455F}

\begin{table}[h!]
\caption{\label{table1}Relative energies (in eV) of the lowest energy structures of Pt$_{5}$V clusters computed at different DFT levels. Two favorable values on the spin multiplicity (S$_M$) are shown. For each functional, the most stable configurations are specified (*).}
\begin{tabular}{llllllll}
{Label}  & S$_M$ & \small{PBE} & \small{PBE0} & \small{PBEh-3c} & \small{B3PW91} & \small{M06-L} & \small{TPSS}\\
\hline
\textbf{5M1.1}$^{b}$ & 2 & 0.27  & 0.06  & 0.48 & 0.05  & 0.49  & 0.28  \\
\textbf{5M1.2}$^{a}$ & 2 & 0.00* & 0.44  & 0.00* & 0.27  & 0.00* & 0.00*  \\
\textbf{5M1.3}       & 2 & 0.04  & 0.57  & 0.84 & 0.33  & 0.19  & 0.14  \\
\textbf{5M1.4}       & 2 & 0.10  & 0.43  & 1.47 & 0.26  & 0.01  & 0.13  \\
\textbf{5M1.5}       & 2 & 0.24  & 0.81  & 0.68 & 0.60  & 0.34  & 0.21  \\
\\
\textbf{5M1.1}       & 4 & 0.45  & 0.00* & 0.53 & 0.00* & 0.29  & 0.32  \\
\textbf{5M1.2}       & 4 & 0.39  & 0.03  & 0.61 & 0.02  & 0.48  & 0.25  \\
\textbf{5M1.3}       & 4 & 0.37  & 0.23  & 0.88 & 0.14  & 0.43  & 0.31  \\
\textbf{5M1.4}       & 4 & 0.41  & 0.63  & 0.76 & 0.49  & 0.36  & 0.41  \\
\textbf{5M1.5}       & 4 & 0.37  & 0.37  & 0.70 & 0.03  & 0.40  & 0.31  \\
\hline
\textsuperscript{a}~Ref. \citenum{JENNINGS201391}\\
\textsuperscript{b}This work.
\end{tabular}
\end{table}

\begin{figure}[ht!]
\begin{center}
\scriptsize
 \includegraphics[scale=0.52]{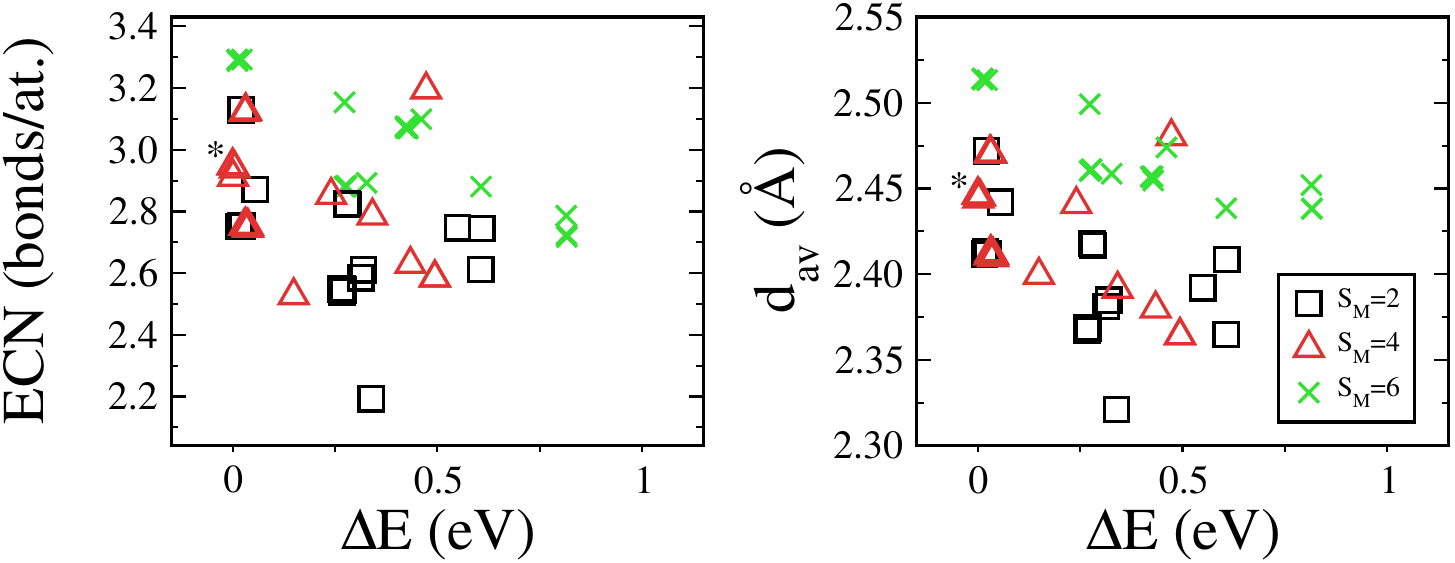}
	\caption{\label{figure_ECN}Distribution of the effective coordination number and average bond distance as a function of the relative energy ($\Delta{E}$) of Pt$_5$V clusters evaluated at the B3PW91/Def2TZVP level. The values on the spin multiplicities (S$_M$) are denoted by distinct symbols. The most stable configuration is specified (*).}
\end{center}
\end{figure}

The results based on these functionals, including the standard PBE, show that two different structures compete for the global minimum (Table~\ref{table1}). The hybrid functionals PBE0 and B3PW91 indicated that the most stable structure is a capped trapezoid ({\bf 5M1.1}) in the quartet state, while the PBE, PBEh-3c, M06-L, and TPSS functionals favor the stability of a pentagonal pyramid ({\bf 5M1.2}) in the doublet state. The later agrees well with the report of Jennings and Johnson.\cite{JENNINGS201391} After analysing the results on various functionals, we further discuss the properties of the most viable structures at the B3PW91/Def2TZVP level, which provides a reasonable approach to describe the bond lengths and ionization energies of Pt clusters.\cite{D3CP04455F} The structural properties are further evaluated using the effective coordination number (ECN) and the average bond distance (d$_{av}$) parameters.\cite{doi:10.1021/acs.jpcc.5b01738} The results show that ECN increases with S$_M$, indicating a dependence on structure and spin (Figure~\ref{figure_ECN}). 

\begin{figure}[h!]
\begin{center}
\scriptsize
 \includegraphics[scale=0.52]{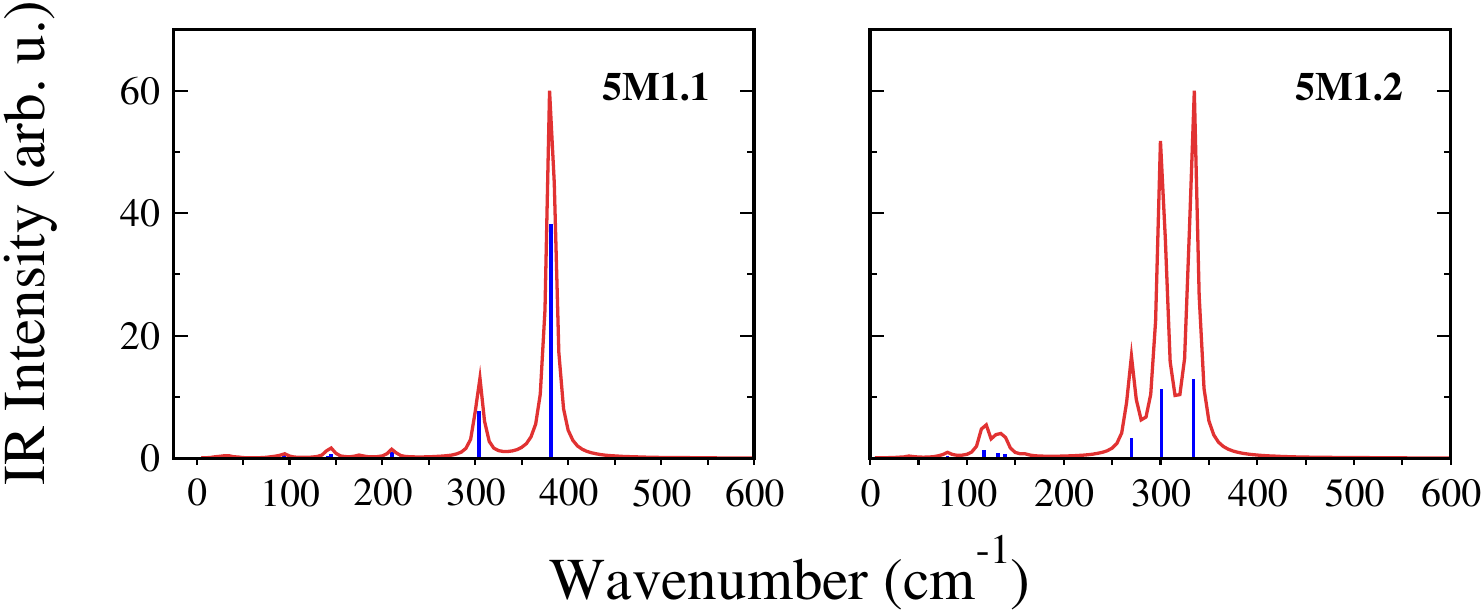}
	\caption{\label{figure_IR}IR spectra for {\bf 5M1.1} and {\bf 5M1.2} clusters obtained at the B3PW91/Def2TZVP level.}
\end{center}
\end{figure}

In order to provide the fingerprints of the most stable structures, we have further calculated the infrared spectra (IR), which will serve as a guide for future experiments when available. The characteristic peaks for {\bf 5M1.1} are found at 304.36 and 381.58 cm$^{-1}$, while for {\bf 5M1.2} they are at 301.23 and 335.16 cm$^{-1}$, respectively (Figure~\ref{figure_IR}). The lowest and highest vibrational frequencies for {\bf 5M1.1} are found at 22.80-381.58 cm$^{-1}$, while for {\bf 5M1.2} are at 29.83-334.16 cm$^{-1}$, respectively, denoting a small range of vibrational spectra. The ionization energy (IP) and electron affinity (EA) parameters are important physical quantities reflecting the electronic stability of clusters.\cite{LAI2021109757,DIE2020113805} The calculated values are shown in Table~\ref{table_1}, while their formulas and definitions are available elsewhere.\cite{RODRIGUEZKESSLER2024122062,RODRIGUEZKESSLER2023116538,RODRIGUEZKESSLER2023121620,doi:10.1021/acs.jpcc.9b03637,MORATOMARQUEZ2020137677} The results show that the Pt$_5$V clusters exhibit only small variations in electronic properties. For the {\bf 5M1.y} (y=1-5) isomers, the IP and EA parameters range from 7.05-7.58 and from 2.16-2.37 eV, respectively. The chemical hardness ($\eta$) and chemical potential ($\mu$) parameters also show minimal variations among these representative isomers. In this context, the differences between the {\bf 5M1.1} and {\bf 5M1.2} clusters are negligible. Moreover, the structures of the clusters exhibited only one rotational axis, consistent with C$_1$ symmetry. The electronic state of the clusters is $^4$A.

	\begin{table}[ht!]
	{
 \caption{\label{table_1}{The symmetry point group, electronic state, ionization energy, electron affinity, chemical hardness and chemical potential of {\bf 5M1.y} (y=1-5) clusters. The results are calculated at the B3PW91 level in conjunction with the Def2TZVP basis set. The energy is given in eV.}}
\centering
\small
\def\arraystretch{1.1}
\begin{tabular}{p{1.7cm}p{1.2cm}p{1.2cm}p{1.2cm}p{1.2cm}p{1.2cm}p{1.2cm}}
	{Cluster} &  Sym  & E$_{state}$	  &  IP     & EA    & $\eta$ & $\mu$    \\ \hline
 {\bf 5M1.1}$^{b}$ & C$_{1}$ & $^4$A  &  7.14  & 2.18  &  2.47  &  4.66  \\  
{\bf 5M1.2}$^{a}$ & C$_{1}$ & $^4$A   &  7.20  & 2.37  &  2.41  &  4.79  \\
 {\bf 5M1.3} & C$_1$ & $^4$A          &  7.58  & 2.57  &  2.50  &  5.07   \\
 {\bf 5M1.4} & C$_1$ & $^4$A          &  7.05  & 2.16  &  2.44  &  4.61   \\
 {\bf 5M1.5} & C$_1$ & $^4$A          &  7.46  & 2.24  &  2.61  &  4.85   \\
 \hline
 \textsuperscript{a}~Ref. \citenum{JENNINGS201391}\\
\textsuperscript{b}This work.
\end{tabular}

	}
\end{table}

\section*{Conclusions}
In this work, we employed density functional theory (DFT) to explore the structure of Pt5V clusters. The results showed that two isomers compete for the global minimum. The distribution of the isomers depends on the spin and coordination of the clusters. The data provided in this preprint serve as a basis for further investigation of the cluster structures as a function of size.


\section*{Acknowledgments}
P.L.R.-K. would like to thank the support of CIMAT Supercomputing Laboratories of Guanajuato and Puerto Interior. 

\nolinenumbers

\bibliography{mendeley}

\bibliographystyle{abbrv}

\section*{Appendix A}


\subsection*{Cartesian coordinates of the Pt$_5$V clusters, obtained by using the Orca program and the B3PW91/Def2TZVP approach. For each cluster, the charge state and spin multiplicity are given.}

{\bf \hspace{0.34cm} 5M1.1} 0,  4\\
 V\hspace{0.51cm}  -0.6954216361045837\hspace{0.51cm}    -0.2379752099514007\hspace{0.51cm}     -0.0105701955035328\\
 Pt\hspace{0.51cm}  0.9553307294845581\hspace{0.51cm}     0.2187780439853668\hspace{0.51cm}      1.8012286424636841\\   
 Pt\hspace{0.51cm} -1.9044703245162964\hspace{0.51cm}     1.5931940078735352\hspace{0.51cm}     -0.5420358180999755\\     
 Pt\hspace{0.51cm} -0.5598647594451904\hspace{0.51cm}    -2.3592939376831055\hspace{0.51cm}     -0.7680397033691406\\   
 Pt\hspace{0.51cm}  1.5922209024429321\hspace{0.51cm}    -1.0426602363586426\hspace{0.51cm}     -0.3284641802310943\\     
 Pt\hspace{0.51cm}  0.6122071146965026\hspace{0.51cm}     1.8279583454132080\hspace{0.51cm}     -0.1521187424659729\\  
  \vspace{0.5cm}
{\bf 5M2.2} 0,  4\\
   V\hspace{0.51cm}   0.01906121550023\hspace{0.51cm}      0.70255818190168  \hspace{0.51cm}    0.33030062648271\\
  Pt \hspace{0.51cm} -0.03429417674258 \hspace{0.51cm}     0.52510394689620 \hspace{0.51cm}    -2.04979137754637\\
  Pt\hspace{0.51cm}  -2.19443188931805 \hspace{0.51cm}     0.57019953060780\hspace{0.51cm}     -0.64572297596758\\
  Pt\hspace{0.51cm}  -1.17169752232956 \hspace{0.51cm}    -0.99780179017499  \hspace{0.51cm}    1.19864273083543\\
  Pt\hspace{0.51cm}   1.36825194704587  \hspace{0.51cm}   -0.61077095761953  \hspace{0.51cm}    1.81312774317462\\
  Pt\hspace{0.51cm}   2.01311042584410  \hspace{0.51cm}   -0.18928891161116 \hspace{0.51cm}    -0.64655674697881\\

\end{document}